# Software Supply Chain Vulnerabilities Detection in Source Code: Performance Comparison between Traditional and Quantum Machine Learning Algorithms

Mst Shapna Akter*, Md Jobair Hossain Faruk*, Nafisa Anjum†, Mohammad Masum§, Hossain Shahriar†,

Akond Rahman¶, Fan Wu††, Alfredo Cuzzocrea‖

*Department of Computer Science, Kennesaw State University, USA
†Department of Information Technology, Kennesaw State University, USA
§Department of Applied Data Science, San Jose State University, USA
¶Department of Computer Science and Software Engineering, Auburn University, USA
††Department of Computer Science, Tuskegee University, USA
‖iDEA Lab, University of Calabria, Rende, Italy

{makter2, mhossa21, nanjum}@students.kennesaw.edu | {hshahria}@kennesaw.edu | mohammad.masum@sjsu.edu |
akond@auburn.edu | fwu@tuskegee.edu |   alfredo.cuzzocrea@unical.it

*Abstract*— **The software supply chain (SSC) attack has become one of the crucial issues that are being increased rapidly with the advancement of the software development domain. In general, SSC attacks execute during the software development processes lead to vulnerabilities in software products targeting downstream customers and even involved stakeholders. Machine Learning approaches are proven in detecting and preventing software security vulnerabilities. Besides, emerging quantum machine learning can be promising in addressing SSC attacks. Considering the distinction between traditional and quantum machine learning, performance could be varies based on the proportions of the experimenting dataset. In this paper, we conduct a comparative analysis between quantum neural networks (QNN) and conventional neural networks (NN) with a software supply chain attack dataset known as ClaMP. Our goal is to distinguish the performance between QNN and NN and to conduct the experiment, we develop two different models for QNN and NN by utilizing Pennylane for quantum and TensorFlow and Keras for traditional respectively. We evaluated the performance of both models with different proportions of the ClaMP dataset to identify the f1 score, recall, precision, and accuracy. We also measure the execution time to check the efficiency of both models. The demonstration result indicates that execution time for QNN is slower than NN with a higher percentage of datasets. Due to recent advancements in QNN, a large level of experiments shall be carried out to understand both models accurately in our future research.**

## I. INTRODUCTION

In recent years, threats to software supply chain security have evolved gradually. Analyzing threat patterns, detecting and predicting security vulnerabilities, and suspicious behaviors of software security threats, Machine Learning (ML) has long been adopted as a powerful approach [1], [2]. Due to the vast level of data stores globally and being enormously increasing by 20% every year, finding innovative approaches to machine learning is needed for proactive prevention and early detection of security threats [3], [4]. Quantum Machine Learning (QML) with the help of quantum random access memory (QRAM) has the potential and scores of research institutions are exploiting the promising QML to deal with large amounts of data [5]–[8]. In general, Quantum Machine Learning refers to an integrated field of quantum computing, quantum algorithms, and classical machine learning where the algorithms are developed to address real-world problems of machine learning [32], [33], leveraging the efficiency and concepts of quantum computing [9], [10].

The fundamental concepts of quantum machine learning including quantum coherence, superposition, and entanglement provide quantum computers with immense power to process and handle data in such a way that leads toward the emerging implementation of quantum computing in technological fields [11], [12]. In contrast to conventional computing, the basic unit of quantum computing known as Qubit, can make use of both the values 0 and 1 in order to follow various paths of computation simultaneously [13]. Mathematically, a qubit state is a vector in two-dimensional space, illustrated by the linear combination of the two basis states ($|0\rangle$, and $|1\rangle$) in a quantum system: $|\psi\rangle = \alpha|0\rangle + \beta|1\rangle$, where $\alpha, \beta \in \mathbb{C}$ are probability amplitudes required to satisfy $|\alpha|^2 + |\beta|^2 = 1$ [14]. Such a sequence of basis states is described as quantum superposition, and correlations between two qubits through a quantum phenomenon are termed entanglement.

With the ever-growing size of data, the average number of sophisticated and complicated cyberattacks and data violations such as software supply chain attacks and network intrusion attacks are also increasing rapidly globally. Software Supply Chain (SSC) attack occurs due to penetration of a vendor's network and insertion of malicious code by a cyber threat actor that jeopardizes the software before the vendor distributes it to the customers [15], [16]. SSC attacks affect the software development, dissemination, and utilization phase and becoming extremely critical due to excessive complications of software development strategies over the years [17]. Such attacks occur during the production phase causing vulnerabilities to downstream consumers. SSC attacks can also disrupt newly developed software through patches or hotfixes or even from the outset, thus compromising the system from the start. Hence, SSC attacks can have a significant negative impact on software users in all sectors by gaining complete control over a software's regular functionality. Hijacking updates, undermining code signing, and compromising open-source code are common techniques exclusively used by threat actors to execute SSC attacks [15], [34].

In Recognition and investigation into SSC attacks, there is an absence of sufficient information concerning mitigating or preventing these risks. On the other hand, a network intrusion attack is an attempt to compromise the security of stored information or data on a computer connected to the network. Two distinct types of activities fall under this definition. First, an attacker can gain unauthorized access to a network, files, or information to steal sensitive data, leaving the data unharmed. An attacker can attempt to gain unauthorized access to user devices, resources, or servers to destabilize the entire network by encrypting, deleting, mishandling, or simply modifying the data [18]. To combat such complex, unlawful and unauthorized attacks, concern grows about preventing attacks from using a quantum machine learning-based paradigm [19]–[21].

In the past years, none to very little research was conducted on the software supply chain vulnerabilities dataset using quantum machine learning perhaps due to the availability of quantum computing resources. However, considering currently available QML-based platforms, Pennylane for instance offers programming quantum computers that enable a new paradigm termed quantum differentiable programming and provides seamless collaboration with other QML tools including IBM quantum, NumPy, and TensorFlow quantum. The main ideology of these applications is flexibility which allows it to know how to make a distinction between various quantum devices and choose the finest algorithm for the task. This paper conducts a comparative analysis between quantum neural networks (QNN) and conventional neural networks (NN) by utilizing a software supply chain attack dataset known as ClaMP. The primary contribution of this research is as follows:

- We adopt both quantum machine learning and conventional machine learning to conduct an experiment on a software supply chain attack dataset.

- We provide a comparative analysis of both QML, and ML's performance based on the findings of the experiments using different proportions of the dataset

We organize the rest of the paper as follows: In Section II, we provide a brief related study on quantum machine learning and traditional machine learning. Section III explains the methodology we adopted for our comparative research. The experimental setting and results are explained in Section IV which includes dataset specification and processing. Section V discusses the findings of this paper. Finally, Section VI concludes the paper.

## II. RELATED WORK

First, Addressing the constraints of traditional machine learning methods, researchers are interested in newly emerging quantum machine learning approach for detecting and preventing software and cybersecurity vulnerability [1]. Various machine learning techniques including neural network, naïve bayes, logistic regression, convolutional neural network (CNN), decision tree and support vector machine are successfully applied for classifying software security activities including malware, ransomware, andnetwork intrusion detection [22]–[24].

Christopher Havenstein *et al.* [25] presented another comparative study based on the performance between Quantum Machine Learning (QML) and Classical Machine Learning (ML). The authors worked on QML algorithms with reproducible code and similar code for ML. Later, quantum variational support vector machines were adopted that show higher accuracy than classical support vector machines. In conclusion, the researchers emphasize the potential of quantum multi-class SVM classifiers for the future.

Luis E. Herrera Rodŕıguez et al. [31] presented a comparative study on various machine learning methods for dissipative quantum dynamics where the authors utilized 22 ML models to predict long-time dynamics of quantum systems. The models include convolutional, and fully connected feed-forward artificial neural networks (ANNs), and kernel ridge regression (KRR).

Mohammad Masum *et al.* [26] conducted research on quantum machine learning (QML) to detect software supply chain attacks [1]. The researchers analyzed speed up the performance of quantum computing by applying scores of novel approaches including quantum support vector machine (QSVM) and quantum neural network (QNN). Utilizing both methods, the authors detected software supply chain attacks in open-source quantum simulators, IBM Qiskit and TensorFlow quantum for instance. According to the research findings, quantum machine learning surpasses classical machine learning in terms of processing speed and computational time.

MJH Faruk *et al.* studied quantum cybersecurity from both threats and opportunity perspectives. The authors have provided a comprehensive review of state-of-the-art quantum computing-based cybersecurity approaches. The research indicated that quantum computing can be utilized to address software security, cybersecurity, and cryptographic-related concerns. On the other hand, the malicious individual also misuses quantum computing against software infrastructure due to the immense power of quantum computers [27].

### III. METHODOLOGY

We adopt Quantum Neural Network (QNN), a subfield of Quantum Machine Learning (QML) for this research and applied the model to the ClaMP dataset. Figure 1 demonstrates the framework representing the implementation process. At first, we pre-processed raw data prior to providing as input to the QML model. We used Python and the shuffle function of the Scikit-Learn (sklearn) library for data preprocessing. We also used the index-reset function and drop function from the python library while labeling the encoder from sklearn library. In order to maintain a balanced number of classes and to avoid imbalanced classes that may lead to the wrong prediction, we ensure the efficiency of all of the separated portions of the dataset created from the ClaMP.

For the experiment, we consider only the balanced portions of the dataset. After applying the shuffle functions, we reset the index towards organizing the dataset in ascending order. The drop function is used to remove columns that are unnecessary and does not contribute to making the prediction. In quantum machine learning models, we need to feed numerical values, so the categorical values were converted into numerical values, and all the numerical values were normalized to maintain a similar scale.

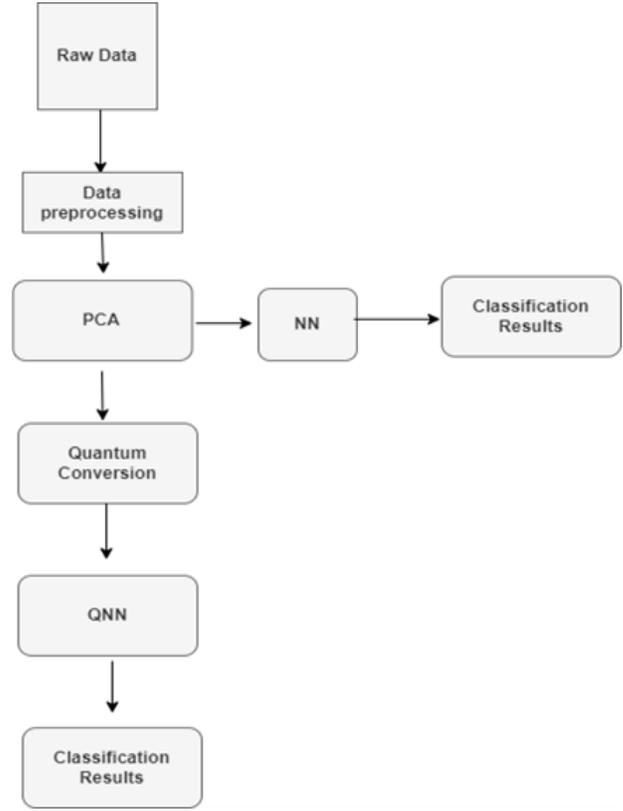

Figure 1: process of the architecture of the framework

After preprocessing steps, we split the entire dataset comprising 5,210 rows into different smaller portions. We separated the dataset into 20 smaller datasets; the number of rows started from 5 percent of the total dataset and gradually increased by 5 percent up to 100 percent. The quantum machine learning model was applied to each of the dataset's separated portions. Before feeding into the QML model, the features were encoded into quantum states. We provide a comparative analysis of both QML and ML's performance based on the findings of the experiments using different proportions of the dataset.

Quantum Neural Network (QNN) comes from neurocomputing theory, which converges with machine learning, quantum computing, and artificial neural network concepts [28]. QNN framework can be applied for processing neural computing utilizing vast levels of datasets to find the expected result. Before processing the data through the QNN, input data is encoded into a suitable qubit state with a proper number of qubits [29]. Later, the qubit state is modified for a specific number of layers using two gates: parameterized rotation gates and entangling gates, where the predicted value of a Hamilton operator, Pauli gates, for instance, is used to direct the altered qubit state. The results derived from Pauli gates are decoded and translated into applicable output data. A variational quantum circuits-based neural network plays various roles in QNN.

Adam optimizer updates the parameters with some criteria including the size of complexity-theoretic measurements, depth, accuracy, and definite features, while the number of steps is necessary for solving the issue of in-depth measurement. Precision describes the setup required to solve a number of challenges. A quantum neural network consists of three items: input, output, and L hidden layers where the L hidden layer consists of a quantum circuit of the quantum perceptron, which acts on an initial state of the input qubits and produces a mixed state for the output qubits. QNN is able to do the quantum computation for also the two input or one input qubit perceptron, which goes through the quantum-circuit construction with quantum perceptron on 4 level qubits. The most comprehensive quantum perceptron implements any quantum channel on the input qubits.

The precision of p(n) is denoted by {s (n), d(n)}, where size is denoted by s(n) and depth is denoted by d(n). The number of qubits in the circuit is measured in size, while the longest sequence of gates from input to output is measured in depth. The size and depth are created from gates D and U of precision p(n). A reversible U gate is usually followed by the D gate to eliminate the localization problem. The accuracy of the circuits is denoted by O{s(n)}.

## IV. EXPERIMENT & RESULTS

In this section, we present both the experiments and results. We first provide details of dataset specification followed by data processing. In order to explain an effective experiment, we define the experimental settings where we utilize accuracy, precision, recall, and F-score metrics for evaluating models' performance. Lastly, we present the experimental results.

### A. Dataset Specification

We applied Quantum Neural Network (QNN) to the ClaMP dataset for malware classification. ClaMP dataset has two versions such as ClaMP_raw and ClaMP_Integrated. The raw instance was aggregated from VirusShare, while the benign instances were integrated from windows files. Portable executable headers contain the information which is required for OS to run executable files. Therefore, features were collected from portable executable headers for malware and benign samples. Moreover, PE header. Hence, various raw features such as File Header (7 features), DOS header (19 features), and Optional header (29 features), were extracted using ruled based method from PE headers of the samples.

Later, the meaningful features are derived using raw features including entropy, compilation time, and section time. Additionally, more information about the PE file was extracted by expanding a set of raw features from the file header. Finally, we selected three types of features including raw, derived, and expanded from the ClaMP_ Integrated dataset, which contains a total of 68 features and the total number of features contains several raw, expanded, and derived features which are 28, 26, and 14 features, respectively [30].

### B. Data Preprocessing

We applied QNN on ClaMP datasets where we utilized various data sizes to inspect the experimented method's comparative performance. We first considered the entire dataset, containing 5210 samples followed by randomly selected 5 percent of the dataset without replacing any instances and gradually increased the percentage by 5 percent, which are 10 percent, 15 percent, 20 percent, 25 percent, 30 percent, 35 percent, 40 percent, 45 percent, 50 percent, 55 percent, 60 percent, 65 percent, 70 percent, 75 percent, 80 percent, 85 percent, 90 percent, and finally 95 percent with 260, 521, 782, 1042, 1302, 1563, 1834, 2084, 2345, 2605, 2566, 3126, 3387, 3647, 3908, 4168, 4429, 4689, and 4950 samples, respectively, while preserving the class proportion. We converted the categorical values from the feature called 'packer type' of ClaMP data since this type of data cannot be directly entered into the model. The dataset contains a total of 108 columns, including one target variable. We used a standardization technique to transform all the features to a mean of zero and a standard deviation of one.

### C. Experimental Settings

The present quantum simulator does not accept large dimensions as input while our dataset contains 108 dimensions, which we cannot feed into the simulator. Hence, we adopted a dimension reduction technique on this dataset called Principal Component Analysis (PCA). PCA was applied to the vector of size 108 features of the CLaMP dataset for reducing the dimension. We selected the first 16 principal components, due to the limitation of qubit numbers in the existing simulator. First, the classical NN was directly applied to the reduced dataset. The next step was encoding the classical data as quantum circuits, which means converting all the features' values into a qubit value for processing in the quantum computer.

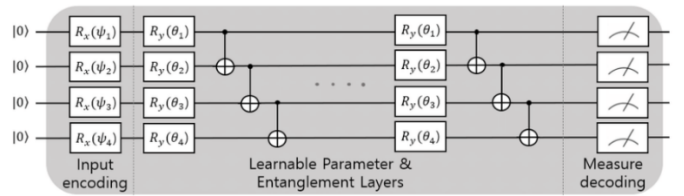

Figure 2: Demonstrates the quantum neural network with the input parameter and Linear entanglement structure

Figure 2 demonstrates the circuit created for a random sample. The circuits were converted into TensorflowQuantum (TFQ). Next, a model circuit layer was developed for the QNN comprising of a two-layer model with the matched size of the data circuit and finally wrapped the model circuit in a TFQ-Keras model. We converted the quantum data and fed it to the model and used a parametrized quantum layer to train the model circuit on the quantum data. We adopt an optimization function known as hinge loss during the training phase. The labels were converted to the -1 to 1 label. Finally, we trained the QNN for 100 epochs.

*D. Experimental Results: Quantum Neural Network (QNN)*

Our comparative analysis between the classical neural network (NN) model and the quantum neural network (QNN) model illustrates in table 1 and Table 2 comprising twenty different portions of the ClaMP dataset. The results derived from the quantum neural network model show that the accuracy is random in the different portions of the dataset. For instance, for 5 percent of the dataset, the accuracy is 57 percent, the f1 score is 73 percent, precision is 100 percent, and recall is 57 percent, while for 20 percent of the dataset, the accuracy is 28 percent, f1 score is 28 percent, recall is 28 percent, and precision is 30 percent. Even if the dataset increases slowly the performance of the model reduces significantly in terms of accuracy. The accuracy suddenly jumps from 30, 35, and 40, to 45 percent and drops off 50 percent of the dataset.

Table 1: displays a comparative analysis of the different portions of the ClaMP dataset using the quantum machine learning model such as QNN

| Data percentage | Precision | Recall | F1-score | Accuracy | Execution Time |
|---|---|---|---|---|---|
| 5 | 1.00 | 0.57 | 0.73 | 0.57 | 12min 24s |
| 10 | 0.42 | 0.35 | 0.37 | 0.35 | 11min 42s |
| 15 | 0.68 | 0.55 | 0.58 | 0.55 | 9min 48s |
| 20 | 0.30 | 0.28 | 0.28 | 0.28 | 9min 22s |
| 25 | 0.64 | 0.47 | 0.53 | 0.48 | 9min 53s |
| 30 | 0.87 | 0.53 | 0.61 | 0.53 | 9min 28s |
| 35 | 0.92 | 0.65 | 0.72 | 0.65 | 10min 23s |
| 40 | 0.89 | 0.60 | 0.65 | 0.60 | 9min 36s |
| 45 | 0.83 | 0.72 | 0.74 | 0.73 | 9min 45s |
| 50 | 0.86 | 0.45 | 0.59 | 0.45 | 9min 30s |
| 55 | 0.86 | 0.50 | 0.63 | 0.50 | 10min 35s |
| 60 | 0.67 | 0.40 | 0.50 | 0.40 | 9min 20s |
| 65 | 0.82 | 0.57 | 0.68 | 0.57 | 9min 45s |
| 70 | 0.80 | 0.50 | 0.62 | 0.50 | 9min 19s |
| 75 | 0.90 | 0.70 | 0.74 | 0.70 | 9min 42s |
| 80 | 0.85 | 0.53 | 0.63 | 0.53 | 9min 40s |
| 85 | 0.82 | 0.45 | 0.51 | 0.45 | 10min 21s |
| 90 | 0.86 | 0.47 | 0.61 | 0.48 | 10min 7s |
| 95 | 0.87 | 0.55 | 0.67 | 0.55 | 10min 24s |
| 100 | 0.93 | 0.53 | 0.67 | 0.53 | 11min 21s |

The accuracy for 30, 35, 40, 45, and 50 percent data are 53, 65, 60, 70, and 45 percent respectively. From the larger portion of the data, 60, 70, 85, and 90, accuracy is incredibly low, which are 40, 50, 45, and 48 percent respectively, while for the data proportion of 65, 75, 80, 95, and 100, accuracy is comparatively high, which are 57, 70, 53, 55, and 53 percent respectively. Considering all of the experiments, the findings indicate that the accuracy is random on different portions of the dataset. The number of instances does not affect the accuracy, while it does affect the total execution time.

Considering the experimental results, the total required execution time is higher when the number of instances is smaller, on the other hand, the execution time starts to decrease when the number of instances increases until a certain threshold. When the data proportion crosses the threshold, the required time gradually starts to increase. Table 1 shows that for 5 percent and 10 percent of the total dataset, the execution times are 12min 24s and 11min 42s respectively. From 15 percent to 80 percent dataset, except from 35 percent and 55 percent, the execution time remains 9 min 19s to 9min 48s. From 85 to 100 percent data, the execution time increases from 10 min to 11 min. Observing quantum neural network models experiment on different portions of data, we found that the performance of the model does not have an effect in terms of accuracy, but the execution time varies with different proportions of the dataset.

*E. Experimental Results: Traditional Neural Network (NN)*

The results derived from the conventional neural network model also show similar results as like the quantum neural network in terms of the accuracy metric, as the accuracy is random in different portions of the dataset. Considering 5 percent of the dataset, the accuracy is 50 percent, the f1 score is 67 percent, the precision is 100 percent, and the recall is 50 percent; for 10 percent of the dataset, the accuracy is 46 percent, the f1 score is 63 percent, the precision is 100 percent, and the recall is 46 percent; for 15 percent of the dataset, the accuracy is 45 percent, the f1 score is 70 percent, the precision is 100 percent, and the recall is 54 percent, which means for the smallest portion of the dataset the accuracy is 50 percent, after increasing the data by 5 percent the accuracy drops by 4 percent, which is 46 percent, and the accuracy increases by 4 percent, which is 54 percent, after increasing the percentage by 10 percent. For the large proportion of the dataset, like 80, 85, 90, and 95 percent, the accuracy values are 53, 52, 54, 54, and 53, respectively.

Table 2: displays a comparative analysis of the different portions of the ClaMP dataset using the classical machine learning model such as NN

| N percent of total data points | Precision | Recall | F1-score | Accuracy | Execution Time |
|---|---|---|---|---|---|
| 5 | 1.00 | 0.50 | 0.67 | 0.50 | 22.1s |
| 10 | 1.00 | 0.46 | 0.63 | 0.46 | 15.2s |
| 15 | 1.00 | 0.54 | 0.70 | 0.54 | 19.8s |
| 20 | 1.00 | 0.48 | 0.65 | 0.48 | 44s |
| 25 | 1.00 | 0.47 | 0.64 | 0.47 | 26.8s |
| 30 | 1.00 | 0.52 | 0.69 | 0.52 | 41.9s |
| 35 | 1.00 | 0.58 | 0.74 | 0.58 | 36.4s |
| 40 | 1.00 | 0.49 | 0.66 | 0.49 | 42.3s |
| 45 | 1.00 | 0.48 | 0.65 | 0.48 | 46.3s |
| 50 | 1.00 | 0.54 | 0.70 | 0.54 | 1min 23s |
| 55 | 1.00 | 0.53 | 0.70 | 0.53 | 1min 22s |
| 60 | 1.00 | 0.51 | 0.68 | 0.51 | 51.1s |
| 65 | 1.00 | 0.55 | 0.71 | 0.55 | 53.6s |
| 70 | 1.00 | 0.55 | 0.71 | 0.55 | 1min 19s |
| 75 | 1.00 | 0.52 | 0.68 | 0.52 | 1min 13s |
| 80 | 1.00 | 0.53 | 0.70 | 0.53 | 1min 16s |
| 85 | 1.00 | 0.52 | 0.69 | 0.52 | 1min 22s |
| 90 | 1.00 | 0.54 | 0.70 | 0.54 | 1min 23s |
| 95 | 1.00 | 0.53 | 0.69 | 0.53 | 1min 17s |
| 100 | 1.00 | 0.52 | 0.68 | 0.53 | 1min 25s |

Analyzing the experimental results, accuracy does not follow any pattern. Neither the accuracy decreases with the different proportion of the dataset, nor does it increase but provides a very unpredictable and random result. Therefore, the model's performance does not have any impact on different proportions of the dataset in terms of the accuracy metrics. However, in terms of the total number of execution times, different portions of the dataset significantly affect the neural network model. We have observed that for smaller portions like 5, 10, 15, 20, 25, 30, 35, 40, and 45 percent of the total dataset the execution times required are 22.1s, 15.2s, 19.8s, 44s, 26.8s, 41.9s, 36.4s, 42.3s, and 46.3s respectively. For 50, 55, 70, 75, 80, 85, 90, 95, 100 percent of the total dataset, the execution times required are 1min 23s, 1min 22s, 1min 19s, 1min 13s, 1min 16s, 1min 22s, 1min 23s, 1min 17s, 1min 25s respectively.

## V. DISCUSSION

The Quantum machine learning model is an emerging approach and has yet to conduct extensive experiments regarding the performance of the proportion of the dataset. In this study, we emphasize experimenting with the quantum machine learning model on different ratios of a dataset and observed how QML works with different ratios of the dataset. Further, we conducted a comparative analysis between the performance of the quantum machine learning model, and the classical machine learning model, to check how the traditional machine learning model works in comparison with the quantum machine learning model.

In accordance with the experiment, QML seems to have a lower influence on various ratios of data in terms of accuracy; however, the efficiency metric is applicable in that case as the efficiency drops with the bigger proportion of the dataset that continues up to a certain limit. The proportion we have chosen are 5, 10, 15, 20, 25, 30, 35, 40, 45, 50, 55, 60, 65, 70, 75, 80, 85, 90, 95, and 100. The accuracy we have found is random, and the execution time decreases from 15 percent of data to 80 percent of data; then again, it starts to increase and continues until 100. The accuracy of the classical machine learning model is also random, but the efficiency starts to drop with the higher ratio of the dataset.

The accuracy, f1 score, precision, and recall results show that the quantum machine learning neural network model and the classical machine learning model does not have an impact on various portions of the dataset. The results show a random pattern throughout the entire dataset. However, we found two specific patterns of execution time results for both models. For the QNN, the execution time decreases with the increment portion of the dataset until a certain threshold of data proportion, for a large number of instances. For the second model, the execution time increases with the increment proportion of the dataset. Therefore, the required execution time is totally opposite of the quantum machine learning model and the classical neural network model using the software vulnerability datasets.

## VI. CONCLUSION

Recently, quantum computing has become a prominent topic with opportunities in the computation of machine learning algorithms that have solved complex problems. This paper conducted a comparative study on quantum neural networks (QNN) and traditional neural networks (NN) and analyzes the performance of both models using software supply chain attack datasets. Due to the limited availability of the quantum computer, the QML model was applied on an open-source Penny lane simulator. We utilized accuracy and processing metrics for evaluating the model's performance. The experimental results indicate that QNN and NN differ in execution time where the QNN model provides quite higher than the NN model. However, the execution time for QNN slows down with the higher proportion of the dataset, while the execution time for NN increases with the higher percentage of the dataset. Although quantum machine learning has been rapidly growing over the last few decades, advancement is still required as the current version of quantum simulators comes with a limited number of qubits, which is not appropriate for software supply chain attacks. A large number of qubits that converges with quantum machine learning models may play a big role in terms of improving classification performance and reducing computation time.

## ACKNOWLEDGEMENT

The work is partially supported by the U.S. National Science Foundation Awards 2209638, 2209636, and 2209637. Any opinions, findings, and conclusions or recommendations expressed in this material are those of the authors and do not necessarily reflect the views of the National Science Foundation.